\documentclass[aps,prl,twocolumn,groupedaddress]{revtex4}

\usepackage{graphicx}
\usepackage{dcolumn}
\usepackage{bm}

\begin{document}

\preprint{APS/123-QED}

\title{Spin-Dependent Quasiparticle Transport in Aluminum Single Electron Transistors}

\author{A. J. Ferguson}
 \email{andrew.ferguson@unsw.edu.au}
\author{S. E. Andresen}%
\author{R. Brenner}%
\author{R. G. Clark}%
\affiliation{%
Australian Research Centre of Excellence for Quantum Computer
Technology, University of New South Wales, Sydney NSW 2052,
Australia
}%

\date{\today}

\begin{abstract}
We investigate the effect of Zeeman-splitting on quasiparticle
transport in normal-superconducting-normal (NSN) aluminum single
electron transistors (SETs). In the above-gap transport the
interplay of Coulomb blockade and Zeeman-splitting leads to
spin-dependence of the sequential tunneling. This creates regimes
where either one or both spin species can tunnel onto or off the
island. At lower biases, spin-dependence of the single quasiparticle
state is studied and operation of the device as a bipolar spin
filter is suggested.
\end{abstract}

\pacs{Valid PACS appear here}
\maketitle

Quasiparticles often feature in the sequential tunneling processes
of nanoscale superconducting devices. Examples include the Josephson
quasiparticle resonances in superconducting single electron
transistors (SETs) \cite{ful89,nak96}, above-gap quasiparticle
Coulomb blockade and even-odd parity effects on superconducting
islands \cite{laf93}. The quasiparticle has spin-$\frac{1}{2}$,
however tunneling rates are not usually dependent on the orientation
of this spin. One exception is in nanoscale aluminum islands with
discrete energy levels where it is possible to directly study the
quasiparticle spin state \cite{ral97}. Spin dependent tunneling of
quasiparticles has been important in large samples where efficient
spin filtering may be performed \cite{mes70,ted71}, and may be of
interest for Coulomb blockaded samples where the behavior of single
quasiparticle spins can be studied.

In this Letter we study the effect of Zeeman-splitting the
quasiparticle states in lithographically fabricated
normal-superconducting-normal (NSN) SETs. In order for Zeeman
splitting to be observed in superconducting films, the effect of the
magnetic field on the quasiparticle orbits must be suppressed. This
is achieved by using ultra-thin (5 nm) aluminum films to confine the
quasiparticle orbits, and precisely aligning the magnetic field in
the plane of the film. This results in a spin-split quasiparticle
density of states first observed by tunneling experiments in
superconducting-normal tunnel junctions \cite{mes70,ted71}. In our
samples the presence of a Zeeman energy (which can be comparable to
both the charging and quasiparticle pairing energies), causes a
difference in energy in creating spin-up and spin-down quasiparticle
excitations on the device island and this leads to several
spin-dependent transport regimes in the above-gap transport. In
addition, quasiparticle populations with a well-defined spin are
created on the island causing accumulation of a magnetic moment. In
the below-gap transport, the Zeeman-splitting changes the energy of
the spin-$\frac{1}{2}$ state of a single quasiparticle on the
island. We observe this energy shift in the sequential tunneling
processes and expect that operation of the device as a bipolar
spin-filter will be possible.

\begin{figure}
\includegraphics[width=7.3cm]{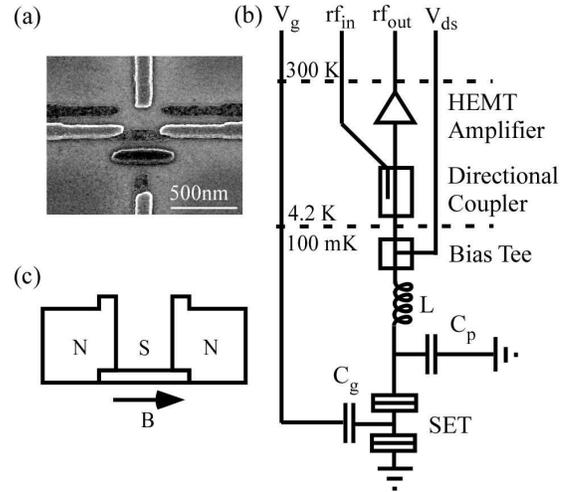}\\
\caption{(a) Micrograph of an SET similar to that measured, showing
a difference in contrast between the 30 nm and 5 nm thick films. To
avoid proximity effects, the normal-state leads do not contact their
superconducting artifacts. (b) Simplified radio frequency circuit
diagram. The $L=470$ nH inductor and parasitic capacitance of
$C_{p}=0.42$ pF form a circuit with a resonant frequency of 321 MHz.
(c) Schematic showing the relative thickness of the device leads and
island. In a magnetic field of $B=500$ mT, the leads are in the
normal state while the island remains superconducting. }\label{}
\end{figure}

The SETs were fabricated by electron beam lithography using a
standard bilayer polymer resist and double angle evaporation
process. The islands were made from 5 nm thick aluminum to allow
Zeeman-splitting, while 30 nm thick aluminum was used for the leads.
In order to achieve electrically continuous 5 nm films, the aluminum
evaporations took place on a liquid nitrogen cooled stage with an
evaporation rate of 0.15 nms$^{-1}$. Between evaporations, the
aluminum was oxidized at an oxygen pressure of 35 mTorr for 5
minutes to form tunnel barriers.

To enable rapid data acquisition and low noise measurement we
configure the devices as rf-SETs \cite{sch98}. The rf-SET technique
involves impedance matching the relatively high SET resistance
towards 50 $\Omega$ by embedding the devices in a resonant circuit
(fig. 1(b)). At resonance, a reflective measurement of a small
incident carrier signal ($V_{rf}\sim$ $\mu$V) is performed, with the
reflected signal detected using a mixer circuit. This experimental
set-up is described in more detail in \cite{bue04}. From the
reflection coefficient and the tank circuit parameters, we can
determine the SET differential conductance. High resistance SETs
(181, 233 k$\Omega$ at 4.2 K) were chosen in order to reduce the
effect of co-tunneling. As a consequence they were poorly impedance
matched by our tank circuit. All measurements were performed in a
dilution refrigerator which achieves an electron temperature of
$\sim100$ mK.

The in-plane critical magnetic field of aluminum is strongly
dependent on film thickness \cite{mes71}. By designing all-aluminum
devices with 30 nm thick leads ($B_c<0.5$ T) and a 5 nm thick island
($B_c=3.8$ T), our superconducting SETs could be turned into
normal-superconducting-normal (NSN) SETs by application of a
magnetic field (fig. 1(a) \& (c)). A limitation of this method is
that we could only approximate a spin-degenerate NSN SET. We
achieved this by applying a $B=500$ mT in-plane magnetic field (fig.
2(a)). At this field, the Zeeman energy ($E_{Z}=g\mu_BB=58$ $\mu$eV)
is greater than the thermal energy ($kT\sim10$ $\mu$eV) but the
transport is not seriously modified from the spin-degenerate case
since $E_{Z}$ is still significantly less than the charging energy
($E_{c}=\frac{e^{2}}{2C_{\Sigma}}=400\mu$eV) and the superconducting
gap ($2\Delta=680$ $\mu$eV).

In NSN SETs the transport characteristics are determined by the
energy required to create quasiparticle excitations, $\Delta$, and
the electrostatic charging energy. The main features are Coulomb
diamonds, due to Coulomb blockade of quasiparticles on the island,
which are offset from zero bias by $V_{ds}=\pm2\Delta$ \cite{eil93,
her94}. Outside the blockaded regions, quasiparticle tunneling
occurs, while in blockade, quasiparticle co-tunneling and low-rate
sequential tunneling processes exist.

\begin{figure}
\includegraphics[width=7.5cm]{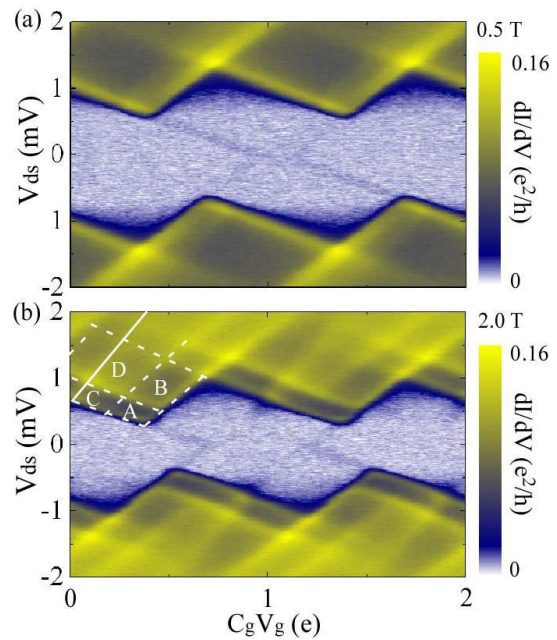}\\
\caption{Measurement of the Coulomb diamonds for the device in NSN
operation as a function of magnetic field. $V_g$ was ramped at 1 kHz
and 256 averages were taken of each trace. (a) At $B=500$ mT the
device approximates an NSN SET. (b) Measurement taken at $B=2$ T.
The Zeeman-split density of states is seen in the above-gap
transport as the emergence of new transport regimes at the onset of
quasiparticle tunneling. The regions A-D between the white dotted
lines correspond to transport regimes described in the text. The
solid white line indicates an unknown spin-dependent transport
process.}\label{}
\end{figure}

As the magnetic field is increased from $B=500$ mT to $B=2$ T, the
Zeeman energy ($E_{Z}=232$ $\mu$eV) becomes comparable to $E_c$ and
$\Delta$. Spin-up and spin-down (spin-down refers to spins
antiparallel to the magnetic field and spin-up to spins parallel)
quasiparticles now have significantly different energies on the
device island. This makes the quasiparticle transport processes
spin-dependent, causing the onset of quasiparticle tunneling to
split (fig. 2(b)) and several new transport regimes (A-D) to emerge.
The allowed transport processes for different bias conditions are
determined by considering the spin-dependent free energy change for
quasiparticles tunneling on and off the island \cite{gandd}.

We consider first quasiparticle transport for the device biased in
region A. When the Coulomb blockade condition permits (determined by
the total charge on the island at a particular time), a
quasiparticle may tunnel from the drain onto spin-down above-gap
states (fig. 3(a)). Above-gap states on the island filled in this
way can also tunnel into the source causing a current to pass. In an
analogous process for below-gap transport, spin-up filled states can
tunnel into the source and quasiparticles from the drain can tunnel
to fill the empty states. For these bias conditions the resultant
current has no net spin-polarization: a spin-down current flows
above-gap and a spin-up current with the same magnitude passes
through the below-gap states.

When biased in region A spin-down quasielectron and spin-up
quasihole populations are generated on the island. Primarily this
occurs since quasiparticle tunneling rates onto and off the island
can be different with, for example, the tunneling rate from the
above-gap filled states depending on the quasiparticle population
\cite{hek93}. A spin-up quasihole has the same magnetic moment as a
spin-down quasielectron and this implies that a magnetic moment
accumulates on the island as a result of the quasiparticle
populations. In the following we estimate this bias dependent
magnetic moment. Approximating to a flat superconducting density of
states and for $V_{ds}-V_T(V_g)>kT$ (where $V_T(V_g)$ is the
conduction threshold including both contributions from the
superconducting gap and Coulomb blockade), the rate of above-gap
quasiparticle tunneling onto the island through a tunnel barrier
with conductance $G$ is $\tau^{-1}\sim\frac{(V_{ds}-V_T)G}{2e}$. By
contrast, the rate for each above-gap quasiparticle to tunnel off
the island is given by $\tau_{esc}^{-1}= \frac{G}{D_{F}e^2}\sim100$
ns \cite{hek93} where $D_{F}=5.8\times10^{6}$ eV$^{-1}$ is the
density of states at the Fermi energy for an island of dimensions 5
nm $\times$ 100 nm $\times$ 500 nm. Equilibrium is reached when the
rates for quasiparticles tunneling on and off the island are equal,
at which point the number of above-gap quasiparticles on the island
is $N_{qe}=\frac{\tau_{esc}}{\tau}$. For $V_{ds}=V_T+100$ $\mu$V
(e.g. approximately in the middle of region A) and the average
measured tunnel barrier conductance of $G=11$ $\mu$S, $N_{qe}=300$.
An equal quasihole population leads to a magnetic moment $M=600$
$\mu_B$ on the island. Quasiparticle recombination reduces these
populations of quasielectrons and quasiholes, however, we expect
that the recombination time is increased from the spin-degenerate
case where $\tau_{r}=1-10$ $\mu$s \cite{lev68,mil67} since a spin
flip is required. Recombination was not considered in the estimate
since $\tau_{r}>\tau_{esc}$.

\begin{figure}
\includegraphics[width=7.3cm]{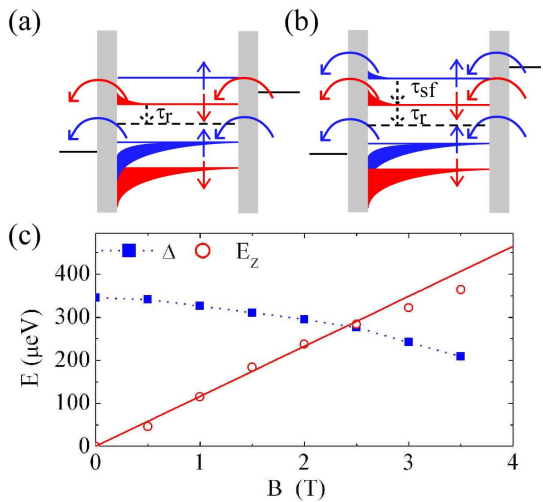}\\
\caption{(a)-(b) Quasiparticle tunneling processes that occur for
the bias conditions A and B in fig. 2(b). Also shown schematically
(dashed arrows) are possible recombination and spin-flip events. (c)
The Zeeman-splitting ($E_{Z}=g\mu_BB$) and mean value of $\Delta$ as
measured from above gap quasiparticle transport. A line fitted
through the Zeeman splitting gives a g-factor of $2.01\pm 0.03$ in
the range $B=0$ - 2.5 T (points at higher fields not included due to
reduced fitting accuracy).}\label{}
\end{figure}

We now discuss some of the features of regime B (fig. 3(b)), noting
that C is similar except that the roles of the above and below-gap
states are reversed. In B quasiparticles may tunnel into both the
spin-up and spin-down above-gap states. However, there is an
asymmetry with respect to the above- and below-gap states, and only
spin-up quasiparticles may tunnel off the island from the below-gap
states. There is an important difference in behavior between
quasiparticles which tunnel into the spin-up and spin-down above-gap
states. For the spin-up case there will be a fast spin-relaxation
due to the availability of empty spin-down states at the same
energy. This is similar to the case in normal state aluminium where
the characteristic spin-flip time is $\tau_{sf}\sim 100$ ps
\cite{jed03}. By contrast, quasiparticles in the spin-down
quasiparticles need to undergo an inelastic process for a spin-flip
which we expect to have a much lower rate.

Finally, for regime D quasiparticles of both polarities are involved
in both above and below-gap transport. However, the spin relaxation
processes apply for the spin-up above and below-gap states as
mentioned above. This relaxation will lead to a spin-imbalance on
the device island. In this respect the quasiparticle transport
behavior differs from that for a conventional SET with a
spin-degenerate density of states.

By fitting the transport thresholds in Coulomb diamonds measured at
magnetic fields from $B=500$ mT to $B=2.5$ T, we determined the
quasiparticle g-factor to be $g=2.01\pm 0.03$. We also extracted
$\Delta$ and found a modest reduction as a function of magnetic
field due to residual orbital effects. The zero field value is
$\Delta=350$ $\mu$eV, which shows a strong enhancement over
$\Delta\sim200$ $\mu$eV for 30 nm films.

There are additional spin-dependent above-gap transport features,
indicated by the solid line in fig. 2(b), that may involve
co-tunneling processes. We do not analyze these features here but
note that a master equation approach would be able to determine the
processes involved \cite{sch94}. All transport characteristics are
reproduced in a second device of similar resistance and charging
energy ($R=233$ k$\Omega$, $E_{c}=475$ $\mu$eV).

We now consider the subgap transport processes near zero bias. If
$E_{c}>\Delta$, as for our samples, an even-odd parity effect occurs
as the system ground state alternates, as a function of gate
voltage, between all Cooper-pairs and a state including a single
quasiparticle \cite{eil93}. In this parity effect the quasiparticle
current is related to a thermal average number of quasiparticles on
the island. Similar energy considerations also cause quasiparticle
poisoning in single Cooper-pair transistors and Cooper-pair boxes.
Close to $V_{ds}=0$, the energy can be written as below. The term
including $p(n)$ takes into account the energy required to create a
quasiparticle excitation of either spin. If $n$, the number of
quasiparticles on the island, is even, then $p(n)=0$, while if $n$
is odd, $p(n)=1$.

\begin{equation}
E=\frac{(ne-C_gV_g)^2}{2C_{\Sigma}}+p(n)(\Delta\pm\frac{1}{2}g\mu_{B}B)
\end{equation}

Figure 4(a) shows the Cooper pair and quasiparticle energy levels
for $B=2$ T as a function of gate bias. Since the spin-degeneracy of
the quasiparticle states is lifted by the Zeeman energy, there are
two parabolas representing the two quasiparticle spin states.
Conductance maxima due to the sequential tunneling of quasiparticles
occur at the degeneracies (fig. 4(b)). Their change in position with
magnetic field reflects a spin-dependence caused by the Zeeman
effect lowering the energy of the spin-down quasiparticle level. We
note that this data, taken with 256 averages and time per trace of
$500$ $\mu$s, could have its signal-to-noise ratio significantly
increased with further averaging and low-pass filtering.

\begin{figure}
\includegraphics[width=7.3cm]{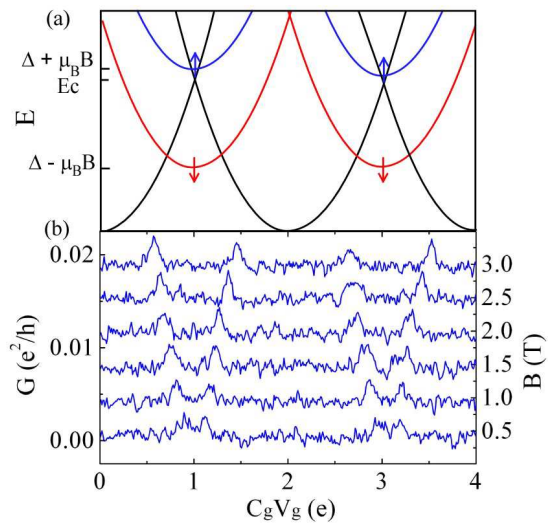}\\
\caption{(a) Energetics of Cooper pair and quasiparticle states at
zero bias for $B=2$ T. The experimental parameters $E_{c}=475$
$\mu$eV, $\Delta (B=2T)=295$ $\mu$eV and $E_z=g\mu_BB=238$ $\mu$eV
are used. The Cooper pair energy is unaffected by the magnetic field
while the quasiparticle levels are split by the Zeeman energy. (b)
Differential conductance (deduced from an averaged rf-measurement)
at $V_{ds}=0$ plotted as a function of B-field. The peaks move apart
as the addition energy of a spin-down quasiparticle is lowered by
the magnetic field.}\label{}
\end{figure}

We now examine the details of these transport processes, first
considering the degeneracy between the 0-Cooper pair level and the
spin-down 1-quasiparticle level. If the system is initially in the
0-Cooper pair state, a spin down quasiparticle may tunnel onto and
then off, the island. This will result in a spin-down current.
Sequential tunneling also occurs when the spin-down 1-quasiparticle
state becomes degenerate with the 1-Cooper pair state. Now, a
spin-up quasiparticle tunnels onto the island forming a singlet
Cooper pair state with the spin-down quasiparticle already on the
island. Subsequently a spin-up tunnels off, breaking a Cooper pair
and leaving a single spin-down quasiparticle on the island. This
gives rise to a spin-up polarized current. Between the degeneracies,
a single above-gap spin-down quasiparticle remains on the island.
The spin-polarization of these currents will be determined by the
rate of spin-flip processes on the island and cotunneling events
involving quasiparticles of the opposite polarity.

The ground state energetics indicate that the device can operate as
a bipolar spin filter. This is analogous to similar behavior
predicted in a GaAs quantum dot \cite{han04,rec00}, where the n = 1
electron Zeeman-split level filters spin-down electrons and the n =
2 singlet state filters spin-up electrons. There is a strong
relationship between these cases since the Cooper pair is a spin
singlet similar to the n=2 electron quantum dot ground state. For
our sample, the measured conductance is $G=0.002\frac{e^{2}}{h}$ and
the peak width $V_{ds}\sim100$ $\mu$V, leading to a maximum current
of $\sim10$ pA. We suggest that it will be possible to obtain
experimental confirmation of bipolar spin filtering by measuring the
current through two independently tunable islands in series.

In conclusion, we have made use of the properties of thin-film
aluminum to Zeeman-split the quasiparticle states on the NSN SET
island. This leads to new spin transport regimes in both above- and
below-gap transport. It will be interesting to further investigate
the spin-filter effects in single and double island devices and
perform studies of quasiparticle spin relaxation and recombination
in these structures.

The authors thank D. J. Reilly, C. M. Marcus, B. J van Wees and F.
E. Hudson for helpful discussions and D. Barber and R. P. Starrett
for technical support. This work is supported by the Australian
Research Council, the Australian government and by the US National
Security Agency (NSA) and US Army Research Office (ARO) under
contract number DAAD19-01-1-0653.

\end{document}